\documentclass[aps,prd,twocolumn,floatfix]{revtex4}
\usepackage{epsfig}

\begin{document}

\title{An Eternal Time Machine in 2+1 Dimensional anti-de Sitter Space}
\author{Simon DeDeo and J. Richard Gott, III}
\affiliation{Department of Astrophysical Sciences, Princeton University, Princeton, New Jersey 08544}

\begin{abstract}
2+1 dimensional anti-de Sitter space has been the subject of much recent investigation. Studies of the behaviour of point particles in this space have given us a greater understanding of the BTZ black hole solutions produced by topological identification of adS isometries. In this paper, we present a new configuration of two orbiting massive point particles that leads to an ``eternal'' time machine, where closed timelike curves fill the entire space. In contrast to previous solutions, this configuration has no event or chronology horizons. Another interesting feature is that there is no lower bound on the relative velocities of the point masses used to construct the time machine; as long as the particles exceed a certain mass threshold, an eternal time machine will be produced.
\end{abstract}

\maketitle

\section{Introduction}

Our time machine is produced by two particles orbiting a common center in circular orbits. In section two, we describe a useful coodinate system in adS. In section three we discuss the nature of the wedge identifications corresponding to the two particles. In section four, we construct the particle orbits and illustrate the resultant spacetime. In section five we examine the causal structure of the solution, and in the final section we compare with previous solutions. 

2+1 dimensional anti-de Sitter space has been the subject of much recent investigation. Holst \cite{hol96} argued that two point masses on nearly radial orbits would create a Gott-type time machine as they passed each other near the center, providing that their relative velocities exceeded a certain mass-dependent threshold related to that for the original Gott configuration in Minkowski space \cite{gott}. Matschull \cite{mat99} noted that two colliding particles would create a BTZ black hole \cite{ban92}. Holst \& Matschull \cite{hm99} found that two light-like sources coming in from infinity and passing each other at a finite distance would create a rotating BTZ black hole with regions of closed timelike curves hidden inside the event horizon, and a ``wormhole'' connection to an additional universe. Many of the results of investigations persuing the connection between single point sources and the BTZ solution are summarized in Birmingham \& Sen \cite{bs00}.

\section{anti-de Sitter space}

Anti-de Sitter space is a homogenous, isotropic space with a negative cosmological constant; 2+1 dimensional anti-de Sitter space can be represented as a three dimensional hyperboloid:

\begin{equation}
\label{surf}
-u^2-v^2+x^2+y^2=-r_0^2
\end{equation}

embedded in a flat four dimensional space with the metric:

\begin{equation}
\label{met}
ds^2=-du^2-dv^2+dx^2+dy^2
\end{equation}

where $r_0^2=-1/\Lambda$, where $\Lambda$ is the cosmological constant as usually defined in the Friedmann equations.

This space is not simply connected and contains closed timelike curves (i.e., the circles $u^2+v^2=w^2\> r_0^2$, with $x$ and $y$ constant and $w^2>1$.) These can be ``unwrapped'' to produce the covering space of anti-de Sitter space which we shall discuss in the remainder of this paper, and simply call anti-de Sitter space. This space has no closed timelike curves.

There are a number of ways to put coordinates on the hyperboloid; we shall be interested in a set that makes the conformal transformation clear. ``Barrel coordinates,'' discussed in Hawking \& Ellis \cite{he73}, chapter $5.2$ and discussed for the 2+1 dimensional case by Holst \& Matschull \cite{hm99}, are the most useful for our situation.

We define $\tau$, a time coordinate, $\chi$, a ``radial'' coordinate, and $\phi$, an angular coordinate as follows:

\begin{eqnarray*}
u & = & r_0\cos\tau\cosh\chi \\
v & = & r_0\sin\tau\cosh\chi \\
x & = & r_0\sinh\chi\cos\phi \\
y & = & r_0\sinh\chi\sin\phi \\
\end{eqnarray*}

the metric induced on the surface is then:

\begin{equation}
ds^2=r_0^2(d\chi^2+\sinh^2\chi\, d\phi^2-\cosh^2\chi\, d\tau^2)
\end{equation}

The time coordinate, $\tau$, runs from $-\infty$ to $\infty$, $\phi$, the angular coordinate is periodic, going from $0$ to $2\pi$, and $\chi$ goes from $0$ to $\infty$. We can set $\tan\theta=\sinh\chi$; since $\chi$ goes from $0$ to $\infty$, $\theta$ goes from $0$ to $\pi/2$. The new metric is then:

\begin{equation}
ds^2=r_0^2\left(\frac{d\theta^2+\sin^2\theta\, d\phi^2-d\tau^2}{\cos^2\theta}\right)
\end{equation}

This demonstrates that anti-de Sitter space is conformally related to the product of a hemisphere and a real time axis. Performing the conformal transformation, and including the equator as a boundary at infinity, we find the ``barrel'' coordinates that we will use in this paper:

\begin{equation}
d\tilde{s}^2=d\theta^2+\sin^2\theta d\phi^2-d\tau^2
\end{equation}

which is related to the original metric by $ds^2=r_0^2\,d\tilde{s}^2/\cos^2\theta$. A property of adS is that null lines starting at the center can reach infinity and return in a finite coordinate time $\Delta\tau=2\pi$.

We note that it is impossible to bring the timelike infinities $\iota^{+}$ and $\iota^{-}$ in to a finite distance without collapsing the spacelike infinities to the origin; since we will only be dealing with processes that elapse over a finite coordinate time, this subtlety will not be an issue.

Each spacelike slice perpendicular to the time axis (i.e., $\tau$ constant) is thus mapped onto a hemisphere. A conformal stereographic projection (with $\rho=\tan\theta/2$) of the half sphere on to the plane:

\begin{equation}
ds^2=\frac{d\rho^2+\rho^2d\phi^2}{(1-\rho^2)^2}
\end{equation}

will project this hemisphere onto a Poincar\'{e} disk of radius $\rho=1$.

\section{Wedge identifications on the Poincar\'{e} disc}

The effect of a point mass in $2+1$ dimensional Minkowski space is to remove a ``wedge'' with angle proportional to the mass of the particle. The two faces of the wedge are identified at constant times in the particle's rest frame. This operation produces a conical singularity at the position of the particle, but leaves the rest of the space flat.

There is an analogous operation for introducing point masses in adS; a general formulation is described in Matschull \cite{mat99}. For our purposes, we need only know how to introduce a stationary point mass at the origin; our spacetimes will be constructed by a procedure of pasting together boosted solutions of this nature.

Our barrel metric gives us a notion of slices of constant time $\tau$ throughout the manifold; as we have seen above, each slice can be conformally mapped onto a Poincar\'{e} disk. The identification procedure is then simple: for a point mass of deficit angle $\alpha$ at the origin $\chi=0$, one identifies two geodesics in the slice of constant time $\tau$, emerging from the point particle, that are separated by $\alpha$ radians of rotation. This produces a conical singularity at the origin. In the limit as one approaches the origin ($\chi<<1$), the curvature term from the cosmological constant is negligible and the geometry near the origin approximates a piece of Minkowski space with a missing wedge of angular size $\alpha$ radians. 

\section{The construction of the eternal time machine}

Our time machine consists of two point masses orbiting a common center in circular orbits. We will construct it by finding the spacetime for a single point mass, boosting into the rest frame of what will be the common center, and putting together two such copies in such a way that the wedges do not intersect.

We begin with a description of the wedge identification lines emerging from single particle at the origin. At each slice of time $\tau$, there is a missing slice of spacetime of angle $\alpha$; the cross section at time $\tau$ looks like a pizza with a slice missing. The location of this missing slice in azimuth is arbitrary. Thus, we may rotate the azimuth of the location of the missing pizza slice at each epoch to produce a ``rotating'' missing wedge whose azimuth is equal to time $\gamma$ and makes one rotation in a period $\gamma=2\pi$. The two edges of the rotating missing wedge may then be described parametrically as follows:

\begin{eqnarray*}
u =  \cos\gamma\cosh\lambda & u^\prime = \cos\gamma\cosh\lambda \\
v = \sin\gamma\cosh\lambda  & v^\prime = \sin\gamma\cosh\lambda \\
x = \cos(\gamma-\alpha/2)\sinh\lambda  & x^\prime = \cos(\gamma+\alpha/2)\sinh\lambda \\
y = \sin(\gamma-\alpha/2)\sinh\lambda  & y^\prime = \sin(\gamma+\alpha/2)\sinh\lambda 
\end{eqnarray*}

\begin{figure}
\epsfig{file=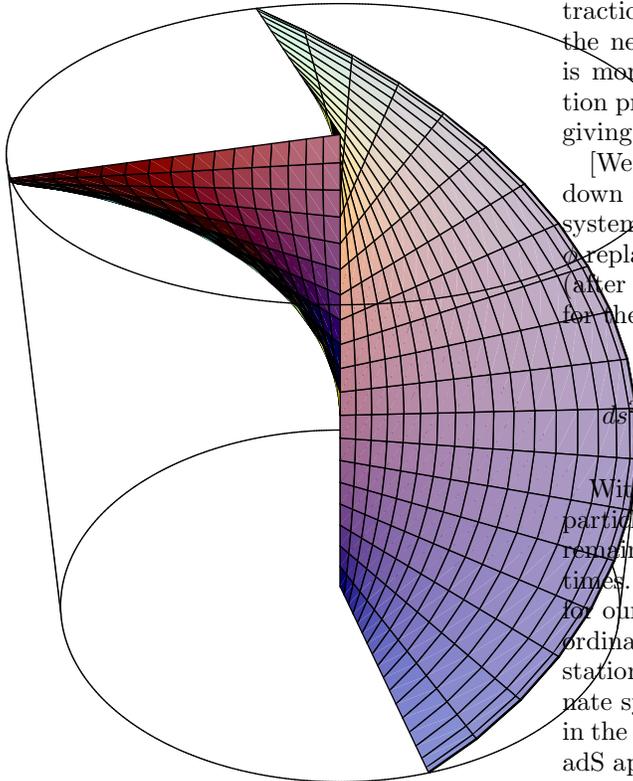,height=120mm}
\caption{The first step of our construction: the wedge associated with the stationary particle at the origin. In this case, the deficit angle $\alpha$ is $\pi/2$. Here we show one half of a period, where the $\tau$ coordinate goes from 0 to $\pi$. For clarity, the wedge identifications are not shown in this figure. We do show, however, the cylinder that is the boundary at infinity. The surface of the cylinder represents $\theta=\pi/2$, corresponding to spatial infinity, while the center of the cylinder is at $\theta=0$. The vertical coordinate represents $\tau$.}
\label{stat}
\end{figure}

where $(u,v,x,y)$ are the coordinates of an event on the trailing edge, and $(u^\prime,v^\prime,x^\prime,y^\prime)$ are the coordinates of the identified event on the leading edge, $\gamma$ is the time coordinate, equal to the proper time of the particle, $\alpha$ is the deficit angle, and $\lambda$, which runs from zero to infinity, parametrizes the distance along the edge, from the particle ($\lambda=0$) to infinity ($\lambda=\infty$). Identifications are to be made between points with the same $\lambda$ and $\gamma$ values. The $\gamma$ dependence in the $x$ and $y$ coordinates is what ``rotates'' the wedges.

We now boost this particle twice, by the same Lorentz boost, $\psi$, but in different directions; the first boost is in the $x-u$ plane, the second in the $y-v$ plane. In the new coordinate system, the equations of the two edges ($\pm$) are:

\begin{eqnarray*}
\label{full}
u & = & \cos\gamma\cosh\lambda\cosh\psi+\cos(\gamma\pm\alpha/2)\sinh\lambda\sinh\psi \\
v & = & \sin\gamma\cosh\lambda\cosh\psi+\sin(\gamma\pm\alpha/2)\sinh\lambda\sinh\psi \\
x & = & \cos(\gamma\pm\alpha/2)\sinh\lambda\cosh\psi+\cos\gamma\cosh\lambda\sinh\psi \\
y & = & \sin(\gamma\pm\alpha/2)\sinh\lambda\cosh\psi+\sin\gamma\cosh\lambda\sinh\psi
\end{eqnarray*}

Again, points with the same $\gamma$ and $\lambda$ are to be identified. After the boosts, the $\gamma$ parameter remains the proper time.

To get a better intution for the behaviour of this spacetime, consider the line parametrized by $\gamma$ with $\lambda=0$ (i.e., the path of the particle itself.) The combination of these two boosts puts the particle into a circular orbit about the new origin of the coordinate system; the particle's worldline is a helix. We note that the proper time of the particle, $\gamma$ is equal to the coordinate time $\tau$:

\begin{eqnarray*}
\phi & = & \arctan(y/x)=\gamma \\
\tau & = & \arctan(v/u)=\gamma
\end{eqnarray*}

The angular position of the particle as a function of $\tau$, in the barrel coordinates, is independent of the boost parameter $\psi$. To build our time machine, we shall introduce a second particle, $\pi$ radians out of phase with the first. The two particles will then helix around each other. The two particles orbit each other eternally without loss of energy through emission of gravity waves because there are no gravitational waves in $2+1$ dimensional spacetime.

Point particles in $(2+1)$ dimensional space exert no gravitational attraction for each other; the particles are kept in circular orbit by the overall gravitational attraction of the negative cosmological constant. (A net attraction is produced because the repulsion produced by the negative mass density of the cosmological constant is more than compensated by the gravitational attraction produced by its positive pressure in two dimensions, giving an overall attraction.)

[We note in passing that, with this insight, we can write down a set of ``corotating'' coordinates, identical to the system described above but with the angular coordinate $\phi$ replaced by $\phi^{\prime}=\phi-\tau$. This gives the following metric (after performing the same conformal transformations as for the original ``barrel'' coordinates):

\begin{equation}
\label{rotation}
ds^2=\rho_0^2\left(\frac{d\theta^2+\sin^2\theta (d\phi^{\prime 2}+2d\phi^{\prime} d\tau)}{\cos^2\theta} -d\tau^2\right)
\end{equation}

With this choice, there is a class of freely falling test particles, orbiting as do our orbiting point masses, that remain at the same radial and azimuthal positions for all times. While this system leads to needless complications for our purposes, it is interesting to note that such a coordinate choice leads one to view the adS manifold as a stationary, rotating universe. There are two such coordinate systems, the second being $\phi^{\prime}=\phi+\tau$, and rotating in the opposite sense. In these global coordinate systems, adS appears as a negatively curved, open, rotating -- but not expanding -- universe.]

\begin{figure}
\epsfig{file=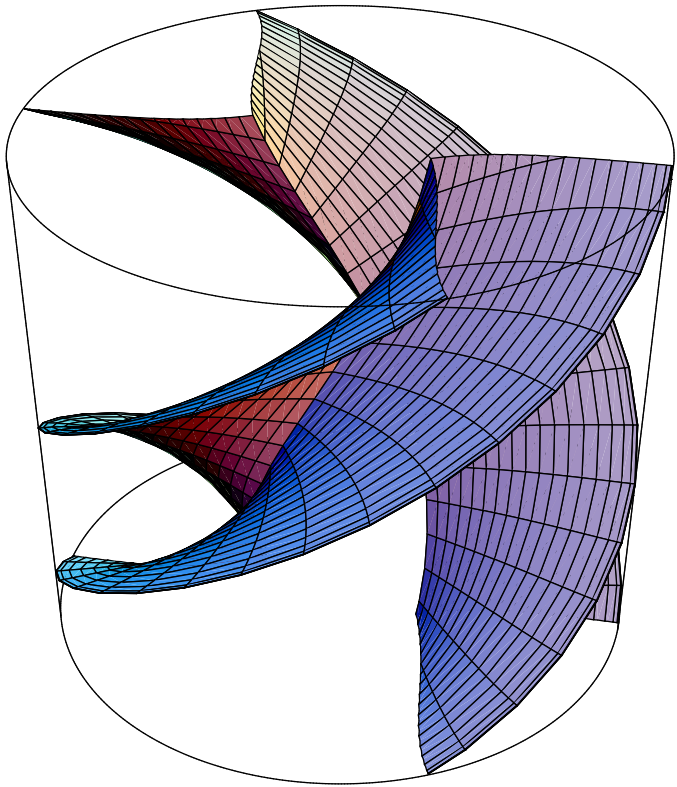,height=120mm}
\caption{The next step in our construction of the time machine. The particle with missing wedge is boosted, and a second particle with missing wedge, $\pi$ radians out of phase, is introduced. As in the previous figure, the deficit angle for each particle, $\alpha$, is $\pi/2$ Here we show one half of a period, where the $\tau$ coordinate goes from 0 to $\pi$. For clarity, the wedge identifications are not shown in this figure.}
\label{spiralwedge}
\end{figure}

As shown in figure \ref{spiralwedge}, the way in which we rotate the wedges in the particle rest frame means that we can do this without worrying about the wedges intersecting in a way that would complicate the simple wedge prescription of the one particle case. Our particluar choice of orientations for the wedges means that the ``leading'' and trailing'' wedge faces are positioned symmetrically about the line emerging from the origin in the direction of each particle.

\section{The causal Structure of the eternal time machine}

We now wish to examine the causal structure of our system. Equation \ref{full}, along with our definitions of the barrel coordinates and our rule for identifying points has given us a full description of the spacetime. We can get a good handle on exactly what is happening by looking at the behaviour of the spacetime at spacelike infinity. In a manner similar to Matschull \cite{mat99}, we will examine the behaviour of null lines that remain always at infinity, and will then be able to make more general statements about the rest of the space.

We can find the identifications at infinity by taking $\lambda$ to infinity in the limit. In this limit, for a mass with deficit angle $\alpha$, the wedge identifications are given by:

\begin{eqnarray}
\label{tau}
\tau & = & \arctan\left(\frac{\sin\gamma+\sin(\gamma\pm\alpha/2)\tanh\psi}{\cos\gamma+\cos(\gamma\pm\alpha/2)\tanh\psi}\right) \\
\label{phi}
\phi & = & \arctan\left(\frac{\sin\gamma\tanh\psi+\sin(\gamma\pm\alpha/2)}{\cos\gamma\tanh\psi+\cos(\gamma\pm\alpha/2)}\right)
\end{eqnarray}

Where the point corresponding to a particular value of $\alpha$, $\psi$, and $\gamma$ is to be identified with the point specified by the same $\psi$ and $\gamma$, and with the $\alpha$ terms having opposite sign. For convenience, we have written these expressions using the arctangent; care must be exercised when dealing with points for which $\phi>\pi/2$ or $\phi<-\pi/2$.

\begin{figure}
\epsfig{file=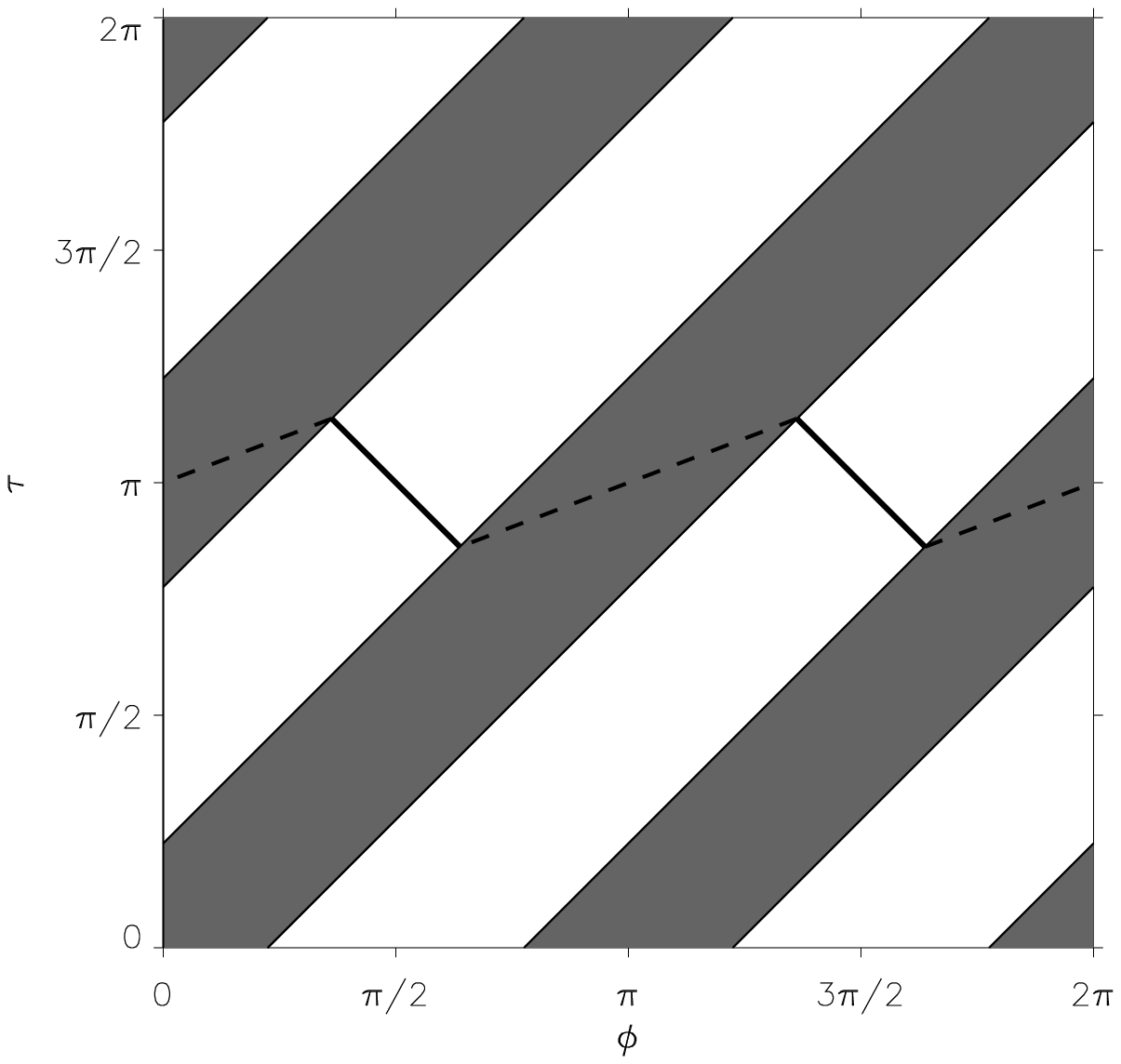,height=120mm}
\caption{The conformal boundary of our spacetime at infinity, unwrapped to make a plane. The time $\tau$ rises on the vertical axis, and $\phi$ goes along the horizontal axis, with $\phi=2\pi$ and $\phi=0$ identified. The shaded regions correspond to the interior of the wedges, which are removed from the spacetime; the edges of these regions are then closed together by identifying according to the dashed lines. The heavy solid lines correspond to the null geodesics orbiting the origin in the opposite sense to the particles; as is shown, the null line closes on itself.}
\label{infinity}
\end{figure}

Following Holst \& Matschull \cite{hm99}, we take the conformal boundary, and, cutting it along $\phi=0$, unwrap it to form a plane. The wedge edges then appear as a parallel lines on this plane. In figure \ref{infinity} we show this construction for $\psi=0.25$ and $\alpha=\pi$; the arrows indicate the identifications. We note that, at infinite distance from the origin, the rotating wedge edges move at a phase velocity equal to the speed of light in order to keep pace with the particle in the interior.

The heavy solid line in this figure shows the path of a light ray at infinity, rotating around the origin in the opposite sense to the two point masses (i.e., clockwise from the $y$ to the $x$ axis.) We see how, for sufficiently large deficit angles, the wedge identifications (depicted by dashed lines) result in the null ray closing on itself. For even larger deficit angles, when the identifications span a larger portion of $(\tau,\phi)$ space, these closed null lines become lines that propagate backwards in time, and one can now draw closed timelike curves on the surface at infinity.

The criteria for closed timelike lines can be found as follows. We require that $\Delta\phi$ and $\Delta\tau$, the amount by which the light ray is identified along its direction of propagation in $\phi$, and the amount by which the light ray is identified backward in $\tau$, respectively, for each wedge are sufficient to recover the time taken by the light ray to circle at infinity. In other words, the light ray can become closed by a combination of being retarded in time ($\Delta\tau$) and by being advanced (clockwise) around the cylinder ($\Delta\phi$) across each wedge identification. If the two wedges taken together thus retard the light ray by a total of $2\pi$, the ray will close. More than that, and closed timelike curves at infinity result. Thus, to make closed timelike curves at infinity we require for each wedge:

\begin{equation}
\label{timetravel}
\Delta\tau+\Delta\phi>\pi
\end{equation}

The expressions for $\Delta\tau$ and $\Delta\phi$ would appear from equations \ref{tau} and \ref{phi} to be rather complicated. Surprisingly, however, they turn out to be very simple. We can examine the case where $\gamma$ is equal to zero (i.e., at the beginning of a cycle); by helical symmetry, our expressions for $\Delta\phi$ and $\Delta\tau$ will be valid for all values of $\gamma$.

For this case, we have

\begin{eqnarray*}
\Delta\tau & = & 2\arctan\left(\frac{\sin(\alpha/2)\tanh\psi}{1+\cos(\alpha/2)\tanh\psi}\right)\\
\Delta\phi & = & 2\arctan\left(\frac{\sin(\alpha/2)}{\tanh\psi+\cos(\alpha/2)}\right)
\end{eqnarray*}

We remind the reader that, as mentioned before, in evaluating these expressions we must keep track of which quadrant we are in (using the signs of the numerator and denominator terms in parentheses) to recover the full range $(0,2\pi)$ of $\Delta\tau$ and $\Delta\phi$, since the arctangent function has only the range $(-\pi/2,\pi/2)$. We can find the expression for the sum $\Delta\tau+\Delta\phi$ by judicious use of the tangent sum formula. We find, after much simplification, for each wedge, the result:

\begin{equation}
\label{difference}
\Delta\tau+\Delta\phi=\alpha
\end{equation}

The dependence on $\psi$ has dropped out. Combining equation \ref{difference} with \ref{timetravel}, we thus find that closed lightlike curves will be produced at infinity by two idenitical orbiting point masses when the sum of their deficit angles is equal to $2\pi$. If the sum exceeds $2\pi$, then closed timelike curves at infinity will result.

Now for point masses in 2+1 dimensional Minkowski space we expect that if the sum of the deficit angles of the two point masses exceeds $2\pi$ (more than a hemisphere of curvature), the spacetime closes like a ``dunce cap'' rather than extending to infinity. This induces additional mass points (at the ``tip'' of the dunce cap) so that the total mass in the closed space equals $4\pi$ as expected from the Gauss-Bonnet theorem in a closed space where the only curvature comes from the point masses. A spacelike section through anti-de Sitter space, however, has uniform negative curvature (it maps onto a Poincar\'{e} disk.) This negative curvature counteracts the positive curvature in the two point masses and allows the spacetime to remain open and extend to infinity even when the sum of the two point masses exceeds $2\pi$.

\begin{figure}
\epsfig{file=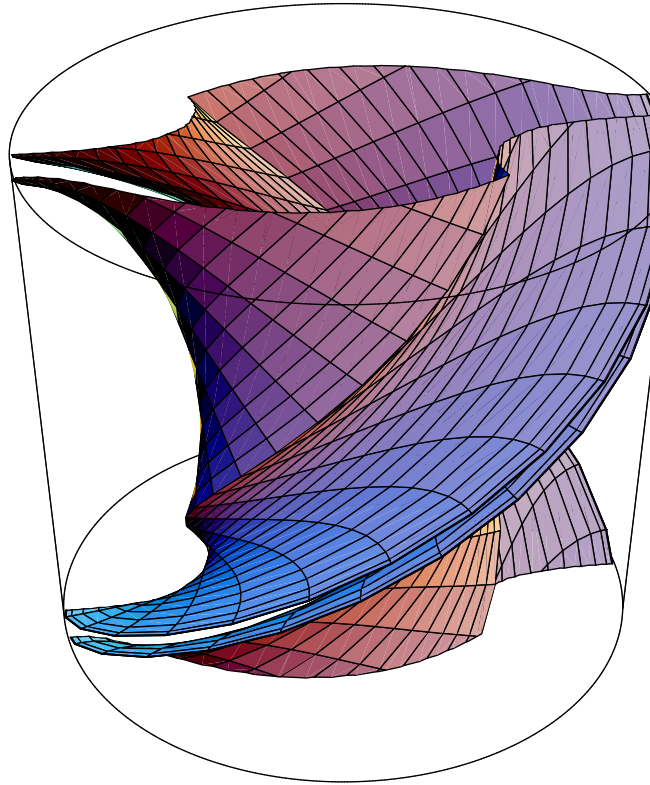,height=120mm}
\caption{A demonstration that the space can support point mass deficit angles greater than $2\pi$ total without closing. Here each point mass has an deficit angle of $3\pi/2$; there are still two small stripes at infinity not removed by the wedge identifications.}
\label{coolwedges}
\end{figure}

So we \emph{are} allowed to have the sum of the point mass deficit angles that sum to greater than $2\pi$ while keeping a subset of points at infinity, by simply separating the two masses by sufficiently large boosts. In figure \ref{coolwedges}, we show this construction (produced with Mathematica) for $\alpha=3\pi/2$ for each point mass, and $\psi=1$. It is clear from this figure that the two wedges do not intersect, and that there are still two ``stripes'' of spacetime at infinity.

We now derive a condition on $\psi$ for the two wedges not to intersect or touch. To derive this condition, it will be convenient to go into a reference frame where one of the particles (the ``stationary particle'') is at the origin. Then, the other particle (the ``orbiting particle'') will orbit at a boost parameter of $2\psi$. The particle at the origin remains stationary, and its wedge, which we will call the ``stationary particle wedge,'' subtends an angle $\alpha$ with identifications at constant $\tau$.

Without loss of generality, we consider the case $\gamma=0$ (all other values of $\gamma$ can be obtained by rotating about the origin), and we remind the reader that our $\phi$ coordinate increases going counterclockwise, starting from $0$ on the line going out from the origin through the orbiting particle at $\gamma=0$.

The lines of identification for the orbiting particle at $\gamma=0$ then start at $\tau=0$, $\phi=0$ at $\lambda=0$,  and run to some maximum $\tau_{\textrm{max}}$ and $\phi_{\textrm{max}}$ as $\lambda$ goes to $\infty$ (we consider, again, without loss of generality because of the reflection symmetry, the leading wedge line only.)

The condition that the wedges never intersect becomes the following: the leading wedge identification line of the orbiting particle may not intersect any part of the stationary particle's wedge in the range $\tau=(0,\tau_{\textrm{max}})$.

The stationary particle's wedge planes, by construction, move around the origin with an angular velocity of unity. For the leading line on the leading edge of the orbiting particle, $\tau(\lambda)$ is always greater than $0$, and increasing monotonically moving out from the orbiting particle (i.e., as $\lambda$ increases.) We can thus write the angular position of the trailing stationary wedge plane as a function of $\tau(\lambda)$, the parameter along the leading line (if the leading line intersects the stationary particle's wedges, it will do so first at the trailing wedge plane.)

We find:

\begin{equation}
\label{trail}
\phi_{\textrm{trail}}=(\pi-\frac{\alpha}{2})+\tau(\lambda)
\end{equation}

which shows the location of the trailing edge of the stationary particle at time $\tau(\lambda)$, where

\begin{equation}
\tau(\lambda)=\arctan\left(\frac{\sin(\alpha/2)}{\coth\lambda\coth\psi+\cos(\alpha/2)}\right)
\end{equation}

Meanwhile, the $\phi$ coordinate of the leading line on the leading wedge plane of the orbiting particle is written:

\begin{equation}
\phi_{\textrm{lead}}=\arctan\left(\frac{\sin(\alpha/2)}{\coth\lambda\tanh\psi+\cos(\alpha/2)}\right)
\end{equation}

Our condition for non-intersection of the wedges then reads:

\begin{equation}
\phi_{\textrm{trail}}>\phi_{\textrm{lead}}
\end{equation}

for all values of $\lambda$ between 0 and $\infty$. In other words, the line parametrized by $\lambda$ for $\gamma=0$ on the leading edge of the orbiting particle should never hit the trailing wedge of the stationary particle.

We can see immediately that in the limit $\psi\rightarrow\infty$, the two arctangent terms become equal, and $\phi_{\textrm{trail}}-\phi_{\textrm{lead}}\rightarrow\pi-\alpha/2$. In other words, we can fit two particles with deficit angles of up to just below $2\pi$ each into our space, provided we separate them by a sufficient boost $\psi$.

For the case $\alpha=\pi$, the bounding condition on the deficit angles to produce a time travel region at infinity, we consider a small boost $\epsilon$. Expanding equation \ref{trail} above to first order in $\epsilon$, we find:

\begin{equation}
\phi_{\textrm{trail}}=2\epsilon\tanh\lambda
\end{equation}

For small $\lambda$, $\phi_{\textrm{lead}}$ goes to $0$ as long as $\epsilon>0$, and we see that $\phi_{\textrm{trail}}-\phi_{\textrm{lead}}\rightarrow\pi/2$, as expected. For large $\lambda$, $\coth\lambda$ is near unity, and we can again expand to first order in $\epsilon$ to find:

\begin{equation}
\phi_{\textrm{trail}}=-2\epsilon\coth\lambda
\end{equation}

Thus, for large $\lambda$,

\begin{equation}
\phi_{\textrm{trail}}-\phi_{\textrm{lead}}\sim 2\epsilon(\coth\lambda+\tanh\lambda)
\end{equation}

and thus a time machine where the masses have deficit angles of $\pi$ each requires only an infinitesimally small boost. This compares to the results of an earlier investigation of closed time-like curves in AdS for particles in near radial orbits \cite{mat99}; there, in a situation similar to that of the original results for closed time-like curves from cosmic strings and point masses in a Minkowski background \cite{gott}, it was found that two equal point masses passing each other on linear trajectories through the origin of AdS produce CTCs when their velocities relative to the center of mass exceed a certain amount:

\begin{equation}
\frac{v}{c}>\cos(\alpha/2)
\end{equation}

where $\alpha\leq\pi$.

So far, we have shown only the existence of closed lightlike curves on the surface at infinity. We now wish to show that this leads to the existence of closed timelike curves throughout the space.

We note that, without the missing wedges, test particles double boosted by $\psi$ orbit the center of the system with period $2\pi$, regardless of their radial distance from the center. Let us denote by $f(\theta)$ the amount, in $\tau$ and $\phi$, that a double boosted particle, rotating at constant distance, is beamed back in time due to the missing wedges, where $\theta$ is the radial coordinate. For deficit angles that sum to greater than $2\pi$, $f(\pi/2)$ will be greater than $2\pi$. Since $f$ is a continuous function of $\theta$, this implies that there are closed timelike curves a finite distance away from the origin (i.e., $f(\theta)>2\pi$ for some $\theta<\pi/2$.) When the deficit angles sum to precisely $2\pi$, there exist only closed null lines at infinity.

Since our solution is stationary (in the rotating coordinate system described in equation \ref{rotation} above), we can then draw a closed timelike curve through any event when the deficit angles sum to more than $2\pi$. We can draw a timelike curve between any two events; the timelike curve simply orbits the two point masses as many times as needed to ``rewind'' sufficiently backwards in $\tau$ to reach the event designated. (Recall that, from any finite distance you can reach the center in a time $\Delta\tau<\pi/2$. So to go from any $(\chi,\phi)$ to any other $(\chi^{\prime},\phi^{\prime})$ requires only a finite $\Delta\tau$.)

\section{Discussion}

We have constructed, in anti-de Sitter space, a system of two particles orbiting about a common center that leads naturally to the existence of closed time-like and null curves. We have derived analytic results on the constraints for the masses and relative separations of the two particles. Closed timelike curves fill the entire spacetime. Timelike curves may be drawn through any two events within the space; our ``time machine'' exists for all times. 

Holst \& Matschull \cite{hm99}, who also considered the effect of point particles in 2+1 dimensional adS, suggested a connection between their solution and the BTZ solution \cite{ban92}. The CTCs in their solution are hidden inside the event horizon of the rotating black hole. The CTCs in our solution, however, are not found in a single region bounded by Cauchy horizons, but penetrate the entire space. In their solution the point particles that generate the CTCs travel on lightlike paths and are ``emitted'' and ``absorbed'' at infinity at a time $\Delta\tau=\pi$ apart. Regions before and after are connected with a BTZ wormhole with a time machine hidden inside. In our case, the particles simply orbit each other for all time, creating an eternal time machine.

SD acknowledges support from an NSF Graduate Research Fellowship and JRG acknowledges support from NSF grant AST-9900772.

\end{document}